\documentclass[english]{article}
\usepackage[T1]{fontenc}
\usepackage[utf8]{inputenc}
\usepackage{geometry}
\usepackage{color}
\usepackage{babel}
\usepackage{float}
\usepackage{amsmath}
\usepackage{amsthm}
\usepackage{setspace}
\usepackage[unicode=true,pdfusetitle,
 bookmarks=true,bookmarksnumbered=false,bookmarksopen=false,
 breaklinks=false,pdfborder={0 0 1},backref=false,colorlinks=false]
 {hyperref}
\begin{document}
\begin{doublespace}
\begin{center}
\textbf{\textcolor{black}{\Large{}Concepts, Components and Collections
of Trading Strategies and Market Color}}{\Large\par}
\par\end{center}

\begin{center}
\textbf{Ravi Kashyap }
\par\end{center}

\begin{center}
\textbf{SolBridge International School of Business / City University
of Hong Kong}
\par\end{center}

\begin{center}
\begin{center}
\today
\par\end{center}
\par\end{center}

\begin{center}
\par\end{center}

\begin{center}
\par\end{center}

\begin{center}
\tableofcontents{}\newpage{}
\par\end{center}
\end{doublespace}
\begin{doublespace}

\section{Abstract}
\end{doublespace}

\begin{doublespace}
\noindent This paper acts as a collection of various trading strategies
and useful pieces of market information that might help to implement
such strategies. This list is meant to be comprehensive (though by
no means exhaustive) and hence we only provide pointers and give further
sources to explore each strategy further. To set the stage for this
exploration, we consider the factors that determine good and bad trades,
the notions of market efficiency, the real prospect amidst the seemingly
high expectations of homogeneous expectations from human beings and
the catch-22 connotations that arise while comprehending the true
meaning of rational investing. We can broadly classify trading ideas
and client market color material into Delta-One and Derivative strategies
since this acts as a natural categorization that depends on the expertise
of the various trading desks that will implement these strategies.
For each strategy, we will have a core idea and we will present different
flavors of this central theme to demonstrate that we can easily cater
to the varying risk appetites, regional preferences, asset management
styles, investment philosophies, liability constraints, investment
horizons, notional trading size, trading frequency and other preferences
of different market participants.
\end{doublespace}
\begin{doublespace}

\section{To Trade or Not to Trade}
\end{doublespace}

\begin{doublespace}
\noindent We can broadly classify trading ideas or strategies (Pardo
2011; Nekrasov 2014; End-note \ref{Trading Strategy}) and market
color (Merkley, Michaely \& Pacelli 2017; Cheng, Liu \& Qian 2006;
Groysberg, Healy \& Chapman 2008; Schipper 1991; Williams, Moyes \&
Park 1996; End-note \ref{Market Color}, \ref{Sell Side}, \ref{Buy Side})
material into Delta-One and Derivative strategies (generally, the
price of Delta one securities have a one to one correspondence with
the price of the underlying: Natenberg 1994; Hull \& Basu 2016; End-notes
\ref{Delta One}, \ref{Derivative Security}), since this acts as
a natural categorization that depends on the expertise of the various
trading desks that will implement these strategies. For each strategy,
we will have a core idea and it is very straight forward to present
different flavors of this central theme that can easily cater to the
varying risk appetites, regional preferences, asset management styles,
investment philosophies, liability constraints, investment horizons,
notional trading size, trading frequency and other preferences of
different market participants. Once we have some market color or a
possible trading idea, the conundrum seen in the title of this section
presents itself, which is whether to implement it and how exactly;
we address some of these concerns through the rest of this article.

\noindent As the number of securities in any trading idea is varied,
the error in the expected returns, the variance of returns, the benefits
of diversification and the liquidity constraints will vary (risk,
return, diversification and liquidity are discussed in: Bodie, Kane
\& Marcus 2014; Elton, Gruber, Brown \& Goetzmann 2009). Hence, it
is important to consider different number of securities as different
variations of the core strategy. The costs of implementing the strategy
will also vary with the number of securities and other parameters
used to create the strategy. Hence, it is important to explicitly
calculate these figures for different variations of the core strategy
(for trading costs see: Perold 1998; Almgren \& Neil 2001; Kashyap
2016b).

\noindent The Execution Risk and the Post Execution Portfolio risk
of the strategy will depend on the Market Impact, Market Risk and
the Timing of the executions that make up the strategy (Kashyap 2016b).
Ways to achieve optimality in this regard are elaborated below in
the “Risk and Best Execution Advisory” point \ref{enu:Risk-and-Best}.
Any investment firm that already has some of these trading strategies,
can look for supplementary information to improve the pre and post
trading experience. Combination strategies (or cross-trading) is possible
when implementing two strategies reduces the overall market exposure
and hence the risk of the blended strategies.  Also, in certain cases,
once we implement one strategy, the additional cost and the risk to
implement another strategy, (marginal cost and risk) is significantly
less and hence it is can be demonstrated that it is beneficial to
implement multiple strategies. 

\noindent These ideas can be implemented mostly with data available
from Bloomberg (Bloomberg 1981; End-note \ref{Bloomberg}) or other
market data vendors. If additional information is required it is mentioned
along with the specific trading idea (we might be able to supplement
any basic data set with specialized data sources).  A statistical
tool like MATLAB (Window 2010; End-notes \ref{Matlab}, \ref{Statistical Software};
a basic database to maintain the time-series data and other information
would be a prudent investment as well) would be required to perform
the regressions and other computations to back-test (Bailey, Borwein,
de Prado \& Zhu 2014) the strategies, estimate the risk and Profit
and Loss (P\&L) levels on these strategies.  For the sake of brevity
and to keep this document simply as an overview of the possibilities,
the computational complexities and implementation details have been
abstracted away from the individual strategy sections below.  As it
will become clear, the design and implementation of these strategies
will require personnel with quantitative skills and training in mathematics
as it applies to finance and economics. In (Kashyap 2018a) we provide
a complete numerical illustration of one trading strategy that can
be beneficial when we expect the market to rebound after a slump or
after a bad economic cycle.

\noindent This paper acts as a collection of various trading strategies
and useful pieces of market information that might help to implement
such strategies. This list is meant to be comprehensive (though by
no means exhaustive) and hence we only provide pointers and give additional
sources to explore each strategy further. To set the stage for this
exploration, we consider the factors that determine good and bad trades,
the notions of market efficiency, the real prospect amidst the seemingly
high expectations of homogeneous expectations from human beings and
the catch-22 connotations that arise while comprehending the true
meaning of rational investing.
\end{doublespace}
\begin{doublespace}

\section{\label{subsec:Characteristics-of-a}Good Ideas, Bad Trades ... Ugly
Risks}
\end{doublespace}

\begin{doublespace}
\noindent We can represent the investment process as a dotted circle
since there is a lot of ambiguity in the various steps involved. The
circle also indicates the repetitive nature of many steps that are
continuously carried out while investing. If we consider the entire
investment management procedure as being akin to connecting the dots
of a circle, then the Circle of Investment can be represented as a
dotted circle with many dots falling approximately on the circumference,
but it hard to have an exact clue about the exact location of the
center or the length of the radius.

\noindent The Equity asset class holds the potential for unlimited
upside and brings with it partial ownership of the firm and hence
some influence over the decision making process. It can be argued
that this premise of boundless profits, coupled with limited losses
or liability and a certain degree of control, make this asset class
an extremely appealing one, contributing to its immense popularity.
Hence, the strategies discussed below are more immediately applicable
to the stocks markets, but it is fairly straight forward to extend
them to other asset classes (for foreign exchange and fixed income
securities, see: Copeland 2008; Tuckman \& Serrat 2011). The various
asset classes can be compared to balloons tethered to the ground,
with the equity balloon having the weakest connections to the ground
and also the weakest controls to guide it, if it is wind-borne. The
lack of a strong controlling factor also makes regime changes much
harder to detect for equities (regime changes are a major shift in
the investment landscape: Wade 2008; Angeletos, Hellwig \& Pavan 2007;
Gasiorowski 1995).

\noindent We need to keep in mind that good traders can make bad trades
and bad traders can make good trades, but over many iterations, the
good traders end up making a greater number of good trades and hence
we provide some distinctions between good and bad trades below (Kashyap
2014b).
\end{doublespace}
\begin{enumerate}
\begin{doublespace}
\item The factors that dictate a good trade or a bad trade depend on the
Time Horizon and the Investment Objective. The time horizon can be
classified into short term, medium term and long term. The investment
objective can be conservative or aggressive. While there are no strict
boundaries between these categories, such a classification helps us
with the analysis and better identification of trades.
\item Any trade that fulfills the investment objective and time horizon
for which it is made is a good trade. Otherwise, it is a bad trade. 
\item On the face of it, we can view good trades as the profitable ones
and bad trades as ones that lose money. But where possible, if we
try and distinguish between proximate causes and ultimate reasons,
it becomes apparent that good trades can lose money and bad trades
can end up making money. 
\item Due to the nature of uncertainty in the social sciences: the noise
around the expected performance of any security; our ignorance of
the true equilibrium; the behavior of other participants; risk constraints
like liquidity, concentration, unfavorable Geo-political events; etc.
implies we would have deviations from our intended results. The larger
the deviation from the intended results, the worse our trade is.
\item What the above implies is that, bad trades show the deficiencies in
our planning (estimation process) and how we have not been able to
take into account factors that can lead our results astray. It is
true that due to the extreme complexity of the financial markets,
the unexpected ends up happening and we can never take into account
everything. We just need to make sure that the unexpected, even if
it does happen, is contained in the harm it can cause. The good thing
about bad trades is the extremely valuable lessons they hold for us,
which takes us through the loop or trials, errors and improvements. 
\item We then need to consider, how a good trade can lose money. When we
make a trade, if we know the extent to which we can lose, when this
loss can occur and that situation ends up happening, our planning
did reveal the possibility and extent of the loss, hence it is a good
trade.
\item The bottom line is that, good trades, or bad trades, are the result
of our ability to come up with possible scenarios and how likely we
think they will happen. The ability to visualize possible outcomes
is related to intelligence. Everything else being equal, someone with
experience or someone who has had the benefit of having participated
in repetitions of similar situations and then having made decisions
in some cases after mistakes in earlier iterations (learning through
trial and error: Ismail 2014; Kashyap 2017; End-note \ref{enu:Taleb and Kahneman discuss Trial and Error / IQ Points})
will be better at facing uncertain situations and solving problems.
\end{doublespace}
\end{enumerate}
\begin{doublespace}
\noindent Zooming back into equities, the following are some other
factors that can contribute to good equity trades. 
\end{doublespace}
\begin{enumerate}
\begin{doublespace}
\item The trade will not soak too much of the available liquidity, as measured
by the average trading volume, unless of course, we wish to take a
controlling stake in the firm (Bernstein 1987; Pástor \& Stambaugh
2003; Bhide 1993; Amihud 2002; Hameed, Kang \& Viswanathan 2010). 
\item It is held by a number of investors. There is more uncertainty if
there are more investors, but it seems to work to our benefit in most
cases. If the number of investors is limited, the possibility of all
of them doing the opposite of what we want is higher and more likely
(Amihud, Mendelson \& Uno 1999; Lakonishok, Shleifer \& Vishny 1992;
Chan \& Lakonishok 1993; Gompers \& Metrick 2001; Asquith, Pathak
\& Ritter 2005; An \& Zhang 2013). 
\item The noise or the randomness is less so that our decisions can be more
accurate. This can be measured by volatility or the price fluctuations
that we see (French, Schwert \& Stambaugh 1987; Schwert 1989; Ang,
Hodrick, Xing \& Zhang 2006; Glosten, Jagannathan \& Runkle 1993). 
\item The firm issuing the securities is not too dependent on any particular
product, profits from a particular region, is not overburdened with
debt, is paying dividends consistently, its price is not too high
compared to its earnings and other fundamental research indicators
(Lancaster 1990; Spence 1976; Bilbiie, Ghironi \& Melitz 2012; Randall
\& Ulrich 2001; Srinivasan, Pauwels, Silva-Risso \& Hanssens 2009). 
\item If we are able to see some pattern in the share price changes, that
is a good trade. This means that this security is exhibiting Non-Markovian
behavior (Turner, Startz \& Nelson 1989; Hassan \& Nath 2005; Cai
1994; Hassan 2009; Litterman 1983). Such behavior is usually hard
to detect, but it comes down to the lens we are using to view the
world or the methods we are using to perform historical analysis (Hamilton
1994; Gujarati 2009; Verbeek 2008).
\item If the security is affected by any asset price bubbles and we are
able to detect the formation of such bubbles (Siegel 2003; Stöckl,
Huber \& Kirchler 2010; Hott 2009; Andersen \& Sornette 2004; Scherbina
\& Schlusche 2014; Kashyap 2016a). 
\item If we are shorting the security and it has a greater tendency for
a downward movement, as exhibited by its skew and other higher moments
(Corrado \& Su 1996; Bakshi, Kapadia \& Madan 2003; Badrinath \& Chatterjee
1991; Badrinath \& Chatterjee 1988; Chen, Hong \& Stein 2001; Barberis,
Mukherjee \& Wang 2016; Amaya, Christoffersen, Jacobs \& Vasquez 2015).
\item So far, we have talked about the unknowns that we know about. What
about the unknowns that we don't know about. The only thing, we know
about these unknown unknowns are that, there must be a lot of them,
hence the need for us to be eternally vigilant (Taleb 2007).
\end{doublespace}
\end{enumerate}
\begin{doublespace}

\subsection{Models of Planes and Seesaws}
\end{doublespace}

\begin{doublespace}
\noindent The analogy of, building a plane and flying it, to constructing
a model and trading with it, will help us consider the associated
risks in a better way. Modeling would be the phase when we are building
a plane, and the outcome of this process is the plane or the model
which we have built; trading would then be the act of flying the plane
in the turbulent skies, which are the financial markets. The modelers
would then be the scientists (also engineers) and the pilots would
be traders. It is somewhat out of the scope of this document to discuss
questions regarding what kind of person can be good at both modeling
and trading. Trading would need a good understanding of what the model
can do and where it will fall short; and building a model will need
to know the conditions under which an model has to operate and the
sudden changes the trading environment might encounter.

\noindent A deterministic world can be made to seem stochastic quite
easily, since randomness is only from the point of the viewer, the
creator of uncertainty (also perhaps, the universe) has no randomness.
Hence, it might be possible to start with a few simple rules that
makes sense intuitively and explain the stochastic behavior of most
phenomenon. Our investment decisions are made over time and so we
set the direction of forward movement in time to be equivalent to
flying the plane forward. Since we cannot see what will happen in
the future; to fly the plane forward, we should not be able to see
what is in front of us. This is equivalent to a plane with the front
windows blackened out. All we have are rear view mirrors (most planes
don't exactly have rear view mirrors, but let us imagine our plane
having one) and windows to the side. 

\noindent As we are cruising along in time, what we have with us is
the historical data or the view from behind and real time data which
is the view from the side, to aid in navigating our way forward or
to the future. We use the historical data to build our model and then
use the data from the present to help us make forecasts for what the
future holds. The modeling would involve using data inputs to come
up with outputs that can help us decide which securities to pick,
or to help set the direction of motion. The trading aspect would involve
using the model outputs and checking if that is the direction in which
we want to be heading, that is actually deciding which securities
to pick, and watching out for cases where the predictions are not
that reliable. 

\noindent We can use multi-factor models (Hamilton 1994; Ng, Engle
\& Rothschild 1992) to decompose overall portfolio risk and help identify
the important sources of risk in the portfolio and link those sources
with aspirations for active return. We need to use the right principles,
the right material and the right processes. 
\end{doublespace}
\begin{enumerate}
\begin{doublespace}
\item The right principles would require understanding certain concepts
that determine the relevant measure of risk for any asset and the
relationship between expected return and risk when markets are tending
towards equilibrium. Examples for these are the Capital Asset Pricing
Model (CAPM), the Arbitrage Pricing Theory or other multi index models
(Sharpe 1964; Ross 1976; Roll \& Ross 1980; Bodie, Kane \& Marcus
2014).
\item The right material translates to having data on the security returns
and choosing the relevant factors. The amount of data and factors
that is available is humongous. We need to use some judgment regarding
how much history to use. We also need to be attuned to significance
and causality among the factors. All this can involve some independent
data analysis (Krejcie \& Morgan 1970; Granger 1981; Adcock 1997;
Hair, Black, Babin, Anderson \& Tatham 1998; MacCallum, Widaman, Zhang
\& Hong 1999; Lenth 2001).
\item The right process would mean using judicious concepts from econometric
/ statistical theory (Bishop, Fienberg \& Holland 2007; Eisenbeis
1977). Some examples would be to check for the stationarity of variables,
to normalize the variables to scale them properly, to see if there
is any correlation between the independent variables and correcting
for it (Multi-Collinearity: Hamilton 1994; Maddala \& Lahiri 1992;
End-note \ref{Multicollinearity}). We need to make sure no variables
that would have an impact are left out (Omitted Variable Bias: Hamilton
1994; End-note \ref{Omitted-variable bias}). 
\end{doublespace}
\end{enumerate}
\begin{doublespace}
\noindent There needs to be a lot of tinkering; this means we need
to have a continuous cycle of coming up with a prototype, testing
how it works and making improvements based on the performance. This
is especially important in the financial markets, since we are chasing
moving targets as the markets have a tendency to be quasi-equilibriums
(we never know what a true equilibrium is but perhaps, the markets
fluctuate between multiple equilibriums, somewhat like a see-saw:
Mantzicopoulos \& Patrick 2010; Stocker 1998; End-note \ref{Seesaw}). 

\noindent Modeling needs to be well thought out, with due regard to
anticipating as many scenarios as possible and building in the relevant
corrective or abortive mechanisms when adverse situations occur. Given
that, we are never close to accomplishing a perfect model, which can
handle all cases without failure and without constant changes, we
would need to constantly supervise the outcomes; hence models that
are simple and robust are better suited, since it is easier to isolate
the points of failure when things get rough. Robust here means producing
similar results under a variety of conditions, with some changes to
the inputs or the controls.
\end{doublespace}
\begin{doublespace}

\subsection{Efficient Assumptions, Rational Myths and Catch-22 Conundrums}
\end{doublespace}

\begin{doublespace}
\noindent A primary question that arises in finance is: whether markets
are efficient? Questions \& Answers (Q\&A) are important. But Definitions
\& Assumptions (D\&A) are perhaps, more important. The good news is
that, Q\&A and D\&A might be in our very DNA, the biological one (Alberts,
Johnson, Lewis, Raff, Roberts \& Walter 2002; End-note \ref{DNA}).
Maybe, DNA hold the lessons from the lives of every ancestor we have
ever had. Evolution is constantly coding the information, compressing
it and passing forward, what is needed to survive better and to thrive,
building what is essential right into our genes (Church, Gao \& Kosuri
2012; Lutz, Ouchi, Liu \& Sawamoto 2013; Kosuri \& Church 2014; Roy,
Meszynska, Laure, Charles, Verchin \& Lutz 2015).

\noindent The assumption made in finance regarding homogeneous expectations
(Levy \& Levy 1996; Chiarella \& He 2001), especially in the derivation
of the efficient frontier, the CAPM and the Capital Market Line (Bodie,
Kane \& Marcus 2014), is stunningly sophisticated, yet seemingly simplistic.
Most people would argue that no two people are alike, so this assumption
does not seem validated (Valsiner 2007; Buss 1985; Plomin \& Daniels
1987). Then again, this assumption is perhaps a very futuristic one,
where we are picking the best habits and characteristics, from our
fellow beings (maybe not just humans?) and at some point in the future,
we might tend to have more in common with each other, fulfilling this
great assumption, which seems more of a prophecy. Again, if we become
too similar then mother nature, or, evolution, will have less to work
with; since more differences tell her which characteristics are better
for certain conditions and many possibilities create stronger survival
potential (Rosenberg 1997; Wilke, ... , \& Adami 2001; Elena \& Lenski
2003; Nei 2013). Too much similarity might not be an issue if survival
is no longer a concern (Kashyap 2018b). But with respect to finance,
we might evolve enough, so that one day, we might have the same expectations
with respect to our monetary concerns. This would also be the day
when the Bid-Offer spread would cease to matter, or, we would be indifferent
to it, making every coffee shop, theater, street corner, pub, or everywhere
… a venue for any product (Kashyap 2015).

\noindent Hence if we vary our definitions of efficiency and the corresponding
assumptions, a useful chain of thoughts and efforts would be to capture,
whether and how much, the actions of various players takes the markets
towards different forms of efficiency (Fama 1970). This implies a
belief that different markets could have different levels and forms
of efficiencies over different times and understanding how the players
are acting could be useful to predict what information could be an
advantage, depending on the efficiency that is believed to be at work.
\end{doublespace}

Perhaps, another way of looking at and understanding this concept
is by pondering over the notion of a rational investor, who has been
defined in multiple ways and extensively discussed (also known in
economic circles as Homo Economicus: Persky 1995; Thaler 2000; End-note
\ref{Homo-Economicus}). Before we consider the Q\&A / D\&A related
to a rational investor, we state a simple game from game theory called
``Guess 2/3 of the average'' (Nagel 1995; another related game is
known as the Keynesian beauty contest: Keynes 2018; Büren, Frank \&
Nagel 2012; Nagel, Bühren \& Frank 2017; End-note \ref{Beauty-Contest}).
The objective of the game is for the participants to guess what 2/3
of the average of the guesses of everyone participating in the game
will be, and where the numbers are restricted to the real numbers
between 0 and 100, inclusive. The winner is the person whose chosen
number is closest to the 2/3 average of all chosen numbers and will
obtain a fixed amount as a payoff that is independent of the stated
number and 2/3. If there is a tie, the prize is divided equally among
all the winners.

There is a unique Nash equilibrium (Nash 1951; Osborne \& Rubinstein
1994; End-note \ref{Nash-Equilibrium}) for this game that can be
found by iterated elimination of weakly dominated strategies. Suppose
that all participants guess the highest possible number, 100, then
clearly the average of all the guesses will be 100. But to win, they
need to guess 2/3 of the average, hence their guess has to be 2/3{*}100,
but if everybody has 2/3{*}100 as their guess, they need to guess
2/3 of 2/3{*}100 or 2/3{*}2/3{*}100 and this process will continue.
Continuing this line of reasoning $k$ times and as $k$ gets larger
and larger, that is as $k\rightarrow\infty$, the guess $g$ of any
participant gets closer and closer to zero. Alternately stated, the
limiting value for the guess game, $g$, as the number of iterations
$k$ increases, (as we continue our line of reasoning by taking the
$2/3$ of the maximum value that someone can guess) is given by, 
\begin{equation}
g=\underset{k\rightarrow\infty}{\lim}\left(\frac{2}{3}\right)^{k}*100=0\label{eq:guess-game}
\end{equation}

This game is usually played to demonstrate that few people, including
students of economics and game theory, actually get the equilibrium
answer of zero (End-note \ref{Guess-Game-Classroom}). Hence if a
rational participant is someone who would want to win this game by
making the right decisions and obtain the winning payoff, it is unclear
whether he should present the above equilibrium guess (Eq \ref{eq:guess-game}).
This example illustrates that other than the non-trivial mathematics
expected from any rational participant, he is supposed to be anticipating
what everyone will do, which could include the possibility that not
everyone participating might be fully rational. Hence his winning
guess might have to be different from the one shown in (Eq \ref{eq:guess-game}),
making him do irrational things unless he can not only read the minds
of everyone but also force his will upon them to be rational. To be
rational he has make sure everyone acts rationally or he has to act
irrationally: a Catch-22 situation (Heller 1999).

In a similar vein, we can define a rational investor as someone who
is not just factoring in the potential investment decisions of every
market participant, but is also having to ensure that everyone is
making the decisions that will be beneficial to him. In the above
game, the decisions of everyone had to be the same, but it could be
different in many other situations, especially in the financial markets
(we will not discuss further the very stimulating topic of whether
and how two different decisions can still be beneficial and rational
to the participants; but it would perhaps be proper material for a
slew of papers and books). The message for us is that either we all
have to become rational investors or nobody can be a rational investor.
Another implication of this chain of thoughts is that if all of us
become rational investors then our expectations might become homogeneous,
making markets efficient and validating the assumptions made to derive
finance and economic theories; though if that happens there might
be no need for economics and finance at that point, since we might
have transcended beyond the need for wealth, wants and other worldly
aspects: another Catch-22 situation.

The computational challenge in this case was limited due to the rather
simplified nature of the exercise. But in other decision making situations
encountered in daily life, the tools and theories of economics prescribe
solving complex optimization problems to arrive at the right results.
If we ponder on the intellectual requirements an agent has to possess
to confront and navigate his daily challenges, it becomes clear that
it is beyond the capabilities of all humans to use such calculations
in their everyday decisions. We want to highlight here that some humans
have created complex but wonderfully elegant solutions for many simple
everyday problems. But even after assuming that such a human would
use the solution he has developed for a particular problem, he would
fail to adopt the solutions created by others for other problems he
encounters since he would have limited expertise with problems in
another domain. 

Clearly, another case is where we might not have developed a solution
ourselves, but we have studied the solutions of others extensively
enough so that we are intimately familiar with it and using it has
become second nature to us. Even if we make the valid argument that
many solutions are related and the marginal effort to master a new
solution is less once someone has good knowledge of a few solutions,
we need to be aware that there are too many new techniques and tools
being introduced and the complexity in solutions is increasing significantly.
This issue of most of us not being familiar with most solutions is
exacerbated with the highly specialized nature of academic research
being conducted and encouraged, coumpounded by the artificial boundaries
we have created by labeling disciplines. Many journals do state that
they encourage multi-disciplinary work but when confronted with work
that truly transcends the fake silos of knowledge we have created,
editors and reviewers struggle to make the right decisions: (Ke, Ferrara,
Radicchi \& Flammini 2015 study how common , “sleeping beauties are
in science”, papers whose relevance has not been recognized for decades,
but then suddenly become highly influential and cited; Gans \& Shepherd
1994 find that journals have rejected many papers that later became
classics; this should make us aware of the possibility that many papers
are being rejected only because their contributions are not being
realized).

We want to emphasize that these are unintended consequences due to
the constraints placed on the actual channeling of research efforts
to knowledge creation and dissemination. One reason why such unwanted
outcomes creep up, despite the fact that journals, editors, scholarly
associations and other research institutions are wonderful innovations
done with the honorable intention of helping us comprehend the cosmos
around us, is because we live in a world that requires around 2000
IQ points to consistently make correct decisions; but the smartest
of us has only a fraction of that (Ismail 2014; Kashyap 2017; End-note
\ref{enu:Taleb and Kahneman discuss Trial and Error / IQ Points}).
Hence, we need to rise, above the urge to ridicule, the seemingly
obvious blunders of others, since without those marvelous mistakes,
the path ahead will not become clearer for us. Someone with 2000 IQ
points is surely a super hero, aptly named IQ-Man. So a rational investor
is this mythical character called IQ-Man, who unlike other super-humans
like Super-Man does not even have a movie about him (for society's
fascination with superheroes or super-humans see: Reynolds 1992).
\begin{doublespace}

\section{Sets of Strategies}
\end{doublespace}
\begin{doublespace}

\subsection{\label{sec:Delta-One-Strategies}Delta-One Strategies }
\end{doublespace}
\begin{enumerate}
\begin{doublespace}
\item \label{enu:Index-Rebalance}Index Re-balance 
\end{doublespace}

\begin{doublespace}
\noindent When an index is re-balanced, certain constituents are removed,
added or their weights in the index are changed. Many participants
try to anticipate these changes and take positions depending on what
they expect to change in any index.  Trading strategies and market
color can be created indicating expected inflow and outflow at a security
level based on the expected weight change or depending on the re-balancing
event (Kostovetsky 2003; Bloom, Gouws \& Holmes 2000; Aked \& Moroz
2015; Chow, Hsu, Kalesnik \& Little 2011).  Basket Trading Ideas to
take advantage of expected inflow and outflow of Index Funds based
on expected weights on the re-balance dates.  We can use Index Tracking
(Beasley, Meade \& Chang 2003; Guastaroba \& Speranza 2012; Dose \&
Cincotti 2005; Stoyan \& Kwon 2010) and Co-Integration (Alexander
1999; Engle \& Clive 1987; Banerjee, Dolado, Galbraith \& Hendry 1993;
Harris 1995; Alexander \& Dimitriu 2005a, b) principles to provide
better estimates of the baskets and capture a certain level of dollar
flows. MSCI (Hau, Massa \& Peress 2009; Chakrabarti, Huang, Jayaraman
\& Lee 2005; End-note \ref{MSCI}) indices are ideal set to start
with and it is possible to extend it to other indices with minor adjustments
depending on the rules used for the re-balancing that could differ
among the many index providers. This strategy would need index membership
information from the relevant index providers. Please see Index Tracking
and Co-Integration in points \ref{enu:Index-Tracking-Baskets}, \ref{enu:Co-Integration-Baskets}
\end{doublespace}
\begin{doublespace}
\item \label{enu:Macro-Theme-Baskets}Macro Theme Baskets 
\end{doublespace}

\begin{doublespace}
\noindent Creation of stock baskets depending on Macro themes (Burstein
1999; Franci, Marschinski \& Matassini 2001; Chen, Leung \& Daouk
2003).  Bounce Basket: Securities expected to benefit the most from
the rebound in the securities markets and the overall economy (Kashyap
2018a).  Short Basket: Securities that are expected to show a significant
fall due to fundamental weakness and an overall drop in the market.
 Securities that perform the best during upward expected Inflationary
moves.  Securities that perform the best during upward movement of
Oil Prices, Gold Prices or other commodities.  Regional baskets that
can best cater to investors looking for regional exposure.  These
baskets are identified by performing a factor analysis (principal
component analysis or regressions can also be used: Hamilton 1994;
End-note \ref{Factor Analysis}) of historical security prices and
other factor indicators (Ludvigson \& Ng 2007; Jolliffe 1986; Shlens
2005; Costello \& Osborne 2005).  We can use Index Tracking and Co-Integration
principles to provide better estimates of the baskets. Please see
Index Tracking and Co-Integration in points \ref{enu:Index-Rebalance},
\ref{enu:Index-Tracking-Baskets}, \ref{enu:Co-Integration-Baskets}.
Macroeconomic data-sources would be required for this set of trading
ideas.
\end{doublespace}
\begin{doublespace}
\item Sector Theme Baskets 
\end{doublespace}

\begin{doublespace}
\noindent These set of strategies are aimed at capitalizing on expected
sector rotations. Depending on various macroeconomic developments
and business cycle trends, different sectors are expected to be either
bullish or bearish. We can create baskets of stocks to benefit most
from these expected trends.  It is possible to put a regional spin
on each of the baskets below.  We can customize this for different
durations of the trades and different notional amounts.  We can use
Index Tracking and Co-Integration principles to provide better estimates
of the baskets. Please see Index Tracking and Co-Integration in points
\ref{enu:Index-Rebalance}, \ref{enu:Index-Tracking-Baskets}, \ref{enu:Co-Integration-Baskets}.
Sector specific data-sources would be required for this set of trading
ideas.

\noindent The following are some of the sector theme baskets. 
\end{doublespace}
\begin{itemize}
\begin{doublespace}
\item Gaming (Casino) Basket 
\item Real Estate Basket 
\item Technology Basket 
\item Financials Basket 
\item Utilities Basket 
\item Travel and Luxury Hotels Basket
\end{doublespace}
\end{itemize}
\begin{doublespace}
\item \label{enu:Index-Tracking-Baskets}Index Tracking Baskets 
\end{doublespace}

\begin{doublespace}
\noindent We can customize this based on two characteristics.  Create
baskets of stocks to track the index with the least tracking error
and a desired number of stocks.  Produce a certain amount of tracking
error and select the number of stocks required to achieve this level
of tracking error. Please see Index Tracking and Co-Integration references
in point \ref{enu:Index-Rebalance}.
\end{doublespace}
\begin{doublespace}
\item \label{enu:Co-Integration-Baskets}Co-Integration Baskets 
\end{doublespace}

\begin{doublespace}
\noindent Pairs Trading Ideas are based on two portfolios (or even
individual securities: Miao 2014) of stocks which have moved together
historically, but are now in diverging positions, so we expect them
to converge back.  This is easily implemented for pairs of securities
in the same sector. Please see Index Tracking and Co-Integration references
in point \ref{enu:Index-Rebalance}.
\end{doublespace}
\begin{doublespace}
\item Portfolio Risk Basket 
\end{doublespace}

\begin{doublespace}
\noindent Many sell side desks take on principal positions (commit
their own capital instead of acting just as an intermediary) on baskets
of stocks that are opposite to client trades.  A basis point quote
is provided to the client based on the market impact to enter the
positions, to exit the positions, volatility of the security prices,
expected duration to enter the position, expected duration to exit
the position and transaction costs.  Inventory Management techniques
(Lancioni \& Howard 1978; Blackstone Jr \& Cox 1985) will need to
be used to manage the overall exposures on the desk that is creating
this strategy and this will be incorporated into the basis point pricing
of the risk basket.
\end{doublespace}
\begin{doublespace}
\item Index Arbitrage Trading Ideas
\end{doublespace}

\begin{doublespace}
\noindent Arbitrage between the prices of index constituents and the
corresponding futures contracts (Chung 1991; Dwyer Jr, Locke \& Yu
1996; Kumar \& Seppi 1994; Nandan, Agrawal \& Jindal 2015; Fung, Lau
\& Tse 2015).  We will need to consider the effect of stock commissions
and futures commissions on the profitability of the trades.  We can
provide early close out options as opposed to close out on the expiration
date.  We will need to consider the market impact of the stock transactions
on the stock price when we put on the trade and the effect of the
same if we do an early close out.
\end{doublespace}
\begin{doublespace}
\item Rich / Cheap Analysis based on Stock Beta
\end{doublespace}

\begin{doublespace}
\noindent We can create baskets of under-priced and overpriced securities
based on the Beta of individual securities versus different indices
acting as a proxy for the market (Black 1992; Isakov 1999; Amihud
\& Mendelson 1989; Fletcher 2000).
\end{doublespace}
\begin{doublespace}
\item Equity Swap Baskets
\end{doublespace}

\begin{doublespace}
\noindent In certain markets, most investors do not have access to
trade the securities and the only way to get access to these markets
is through swaps from a registered broker (Kijima \& Muromachi 2001;
Gay, Venkateswaran, Kolb \& Overdahl 2008; Chance \& Rich 1998; Wang
\& Liao 2003; Wu \& Chen 2007; End-note \ref{Equity Swap}). We can
create swap baskets to pick up exposure for the above themes; also
swap baskets can be created for the other markets as well, since it
could be a good alternative to trading all the other investments.
\end{doublespace}
\begin{doublespace}
\item \label{enu:Risk-and-Best}Risk and Best Execution Metrics
\end{doublespace}

\begin{doublespace}
\noindent We can create risk metrics to supplement the information
on trade execution in terms of market impact, market risk and optimal
timing of transactions (Smithson 1998; Bouchaud \& Potters 2003; Wipplinger
2007; Christoffersen 2012). It would be helpful to collect information
on the market impact of individual transactions or portfolios based
on the timing requirements of implementing the strategies. This market
impact can be calibrated to the orders, executions and the skill of
the executions teams within the firm so that it will capitalize on
the trading advantages possessed by the firm.  Such information can
not only help with the planning of optimal execution strategies to
reduce the market impact and market risk of the resulting portfolio,
but can be useful to check if the trading strategy is viable; since
even though it promises to produce good returns but implementing it
might eat away the returns. Optimal execution will be in terms of
the number of pieces the overall basket will be broken into and the
duration and timing over which to trade the individual pieces.  Portfolio
risk monitoring will be needed for the individual strategies and changes
in risk levels will need to be monitored for different increments
/ decrements to the various trading strategies.
\end{doublespace}
\end{enumerate}
\begin{doublespace}

\subsection{Derivative Strategies }
\end{doublespace}

\begin{doublespace}
\noindent To implement these, we would need some derivatives pricing
and risk modules, a basic level of which can be implemented in MATLAB.
Since the number of combinations of derivative instruments is huge,
only a brief overview is provided below. Depending on the specifics
of the instruments and the market conditions, various strategies can
be implemented.
\end{doublespace}
\begin{enumerate}
\begin{doublespace}
\item We can have trades on index options and options on the constituents
based on the volatility of the individual securities and the volatility
level of the index and the correlation between individual security
pairs and the average level of correlation in the basket. These are
known as volatility and dispersion Trades (Marshall 2009; Meissner
2016; Driessen, Maenhout \& Vilkov 2009); 
\item We predict volatility levels using ARCH / GARCH models and trade instruments
that are theoretically mispriced when compared to the implied volatility
levels (Andersen, Bollerslev, Diebold \& Labys 2003; Andersen \& Bollerslev
1998; Engle \& Ng 1993; Xu \& Malkiel 2003; Bollerslev, Chou \& Kroner
1992; Nelson 1991). 
\item Put – Call Parity violations (Cremers \& Weinbaumv 2010; Klemkosky
\& Resnick 1979; Finucane 1991), combined with either implied volatilities
or based on predicted ARCH / GARCH volatility levels, can be used
to find undervalued or overvalued instruments and appropriate strategies
can be constructed. 
\item Option hedges are possible for basket trades constructed in the delta
one section \ref{sec:Delta-One-Strategies}. 
\item Structures based on different options, that is structured equity derivatives
(Kat 2001) suitable for different Macro, Sector themes and Expectations.
\end{doublespace}
\end{enumerate}
\begin{doublespace}

\section{Conclusion}
\end{doublespace}

\begin{doublespace}
\noindent We have looked at a number of delta-one and derivative trading
strategies, related market color and pointers to practically implement
these investment ideas. The advantages of investing in a moderate
amount of computing infrastructure and hiring personnel with technical
abilities were illustrated in terms of having a wider choice of investment
alternatives, increased returns and better management of risky outcomes.
In Kashyap (2018) we discuss in detail a trading strategy, called
the bounce basket, for someone to express a bullish view on the market
by allowing them to take long positions on securities that would benefit
the most from a rally in the markets. Given the dynamic nature of
the financial markets, it would be practical to have a feedback loop
that changes the parameters of any trading strategy depending on changes
in market conditions (Kashyap 2014a).
\end{doublespace}
\begin{doublespace}

\section{End-notes}
\end{doublespace}
\begin{enumerate}
\begin{doublespace}
\item \label{Trading Strategy}\href{https://en.wikipedia.org/wiki/Trading_strategy}{Trading Strategy, Wikipedia Link}
In finance, a trading strategy is a fixed plan that is designed to
achieve a profitable return by going long or short in markets. The
main reasons that a properly researched trading strategy helps are
its verifiability, quantifiability, consistency, and objectivity.
For every trading strategy one needs to define assets to trade, entry/exit
points and money management rules. Bad money management can make a
potentially profitable strategy unprofitable.
\item \label{Market Color} Market Color is a word commonly used on trading
desks both on the buy side and sell side (End-notes \ref{Sell Side},
\ref{Buy Side}). It refers to information regarding the financial
markets, many times changes in variables that are deemed relevant
to make trading decisions. The sell side provides many such pieces
of information to the buy side, eventually hoping to get trades done
on the back of this such material. This could also be considered as
research done by analysts on both the sell side and the buy side.
\item \label{Sell Side}\href{https://en.wikipedia.org/wiki/Sell_side}{Sell Side, Wikipedia Link}
Sell side is a term used in the financial services industry. The three
main markets for this selling are the stock, bond, and foreign exchange
market. It is a general term that indicates a firm that sells investment
services to asset management firms, typically referred to as the buy
side, or corporate entities. One important note, the sell side and
the buy side work hand in hand and each side could not exist without
the other. These services encompass a broad range of activities, including
broking/dealing, investment banking, advisory functions, and investment
research.
\item \label{Buy Side}\href{https://en.wikipedia.org/wiki/Buy_side}{Buy Side, Wikipedia Link}
Buy-side is a term used in investment firms to refer to advising institutions
concerned with buying investment services. Private equity funds, mutual
funds, life insurance companies, unit trusts, hedge funds, and pension
funds are the most common types of buy side entities. In sales and
trading, the split between the buy side and sell side should be viewed
from the perspective of securities exchange services. The investing
community must use those services to trade securities. The \textquotedbl Buy
Side\textquotedbl{} are the buyers of those services; the \textquotedbl Sell
Side\textquotedbl , also called \textquotedbl prime brokers\textquotedbl ,
are the sellers of those services.
\item \label{Delta One}\href{https://en.wikipedia.org/wiki/Delta_One}{Delta One, Wikipedia Link}
Delta One products are financial derivatives that have no optionality
and as such have a delta of (or very close to) one – meaning that
for a given instantaneous move in the price of the underlying asset
there is expected to be an identical move in the price of the derivative.
Delta one products can sometimes be synthetically assembled by combining
options. 
\item \label{Derivative Security}\href{https://en.wikipedia.org/wiki/Derivative_(finance)}{Derivative (finance), Wikipedia Link}
In finance, a derivative is a contract that derives its value from
the performance of an underlying entity. This underlying entity can
be an asset, index, or interest rate, and is often simply called the
\textquotedbl underlying.\textquotedbl{} Derivatives can be used
for a number of purposes, including insuring against price movements
(hedging), increasing exposure to price movements for speculation
or getting access to otherwise hard-to-trade assets or markets.
\item \label{Bloomberg}\href{https://en.wikipedia.org/wiki/Bloomberg_L.P.}{Bloomberg, Wikipedia Link}
Bloomberg L.P. is a privately held financial, software, data, and
media company headquartered in Midtown Manhattan, New York City. It
was founded by Michael Bloomberg in 1981, with the help of Thomas
Secunda, Duncan MacMillan, Charles Zegar, and a 30\% ownership investment
by Merrill Lynch.
\item \label{Matlab}\href{https://en.wikipedia.org/wiki/MATLAB}{Matlab, Wikipedia Link}
MATLAB (matrix laboratory) is a multi-paradigm numerical computing
environment and proprietary programming language developed by MathWorks.
MATLAB allows matrix manipulations, plotting of functions and data,
implementation of algorithms, creation of user interfaces, and interfacing
with programs written in other languages, including C, C++, C\#, Java,
Fortran and Python.
\item \label{Statistical Software}\href{https://en.wikipedia.org/wiki/List_of_statistical_packages}{Statistical Software, Wikipedia Link}
Statistical software are specialized computer programs for analysis
in statistics and econometrics.
\item \label{enu:Taleb and Kahneman discuss Trial and Error / IQ Points}\href{https://www.youtube.com/watch?v=MMBclvY_EMA}{Nassim Taleb and Daniel Kahneman discuss Trial and Error / IQ Points, among other things, at the New York Public Library on Feb 5, 2013.}
As Taleb explains, ``it is trial with small errors that are important
for progress''. The emphasis on small errors is especially true in
a portfolio management context, since a huge error could lead to a
blow up of the investment fund. (Ismail 2014) mentions the following
quote from Taleb, “Knowledge gives you a little bit of an edge, but
tinkering (trial and error) is the equivalent of 1,000 IQ points.
It is tinkering that allowed the industrial revolution''.
\item \label{Multicollinearity}\href{https://en.wikipedia.org/wiki/Multicollinearity}{Multicollinearity, Wikipedia Link}
In statistics, multicollinearity (also collinearity) is a phenomenon
in which one predictor variable in a multiple regression model can
be linearly predicted from the others with a substantial degree of
accuracy. In this situation the coefficient estimates of the multiple
regression may change erratically in response to small changes in
the model or the data. Multicollinearity does not reduce the predictive
power or reliability of the model as a whole, at least within the
sample data set; it only affects calculations regarding individual
predictors. That is, a multivariate regression model with collinear
predictors can indicate how well the entire bundle of predictors predicts
the outcome variable, but it may not give valid results about any
individual predictor, or about which predictors are redundant with
respect to others.
\item \label{Omitted-variable bias}\href{https://en.wikipedia.org/wiki/Omitted-variable_bias}{Omitted-variable bias, Wikipedia Link}
In statistics, omitted-variable bias (OVB) occurs when a statistical
model leaves out one or more relevant variables. The bias results
in the model attributing the effect of the missing variables to the
estimated effects of the included variables. More specifically, OVB
is the bias that appears in the estimates of parameters in a regression
analysis, when the assumed specification is incorrect in that it omits
an independent variable that is correlated with both the dependent
variable and one or more of the included independent variables.
\item \label{Seesaw}A seesaw (also known as a teeter-totter or teeterboard)
is a long, narrow board supported by a single pivot point, most commonly
located at the midpoint between both ends; as one end goes up, the
other goes down. \href{https://en.wikipedia.org/wiki/Seesaw}{Seesaw, Wikipedia Link}
\item \label{DNA}\href{https://en.wikipedia.org/wiki/DNA}{DNA, Wikipedia Link}
Deoxyribonucleic acid (DNA) is a molecule composed of two chains that
coil around each other to form a double helix carrying the genetic
instructions used in the growth, development, functioning, and reproduction
of all known living organisms and many viruses. DNA and ribonucleic
acid (RNA) are nucleic acids; alongside proteins, lipids and complex
carbohydrates (polysaccharides), nucleic acids are one of the four
major types of macromolecules that are essential for all known forms
of life.
\item \label{Homo-Economicus}\href{https://en.wikipedia.org/wiki/Homo_economicus}{Homo Economicus, Wikipedia Link}
The term homo economicus, or economic man, is a caricature of economic
theory framed as a \textquotedbl mythical species\textquotedbl{}
or word play on homo sapiens, and used in pedagogy. It stands for
a portrayal of humans as agents who are consistently rational and
narrowly self-interested, and who usually pursue their subjectively-defined
ends optimally.
\item \label{Beauty-Contest}\href{https://en.wikipedia.org/wiki/Keynesian_beauty_contest}{Keynesian Beauty Contest, Wikipedia Link}.
Keynes described the action of rational agents in a market using an
analogy based on a fictional newspaper contest, in which entrants
are asked to choose the six most attractive faces from a hundred photographs.
Those who picked the most popular faces are then eligible for a prize.
A naive strategy would be to choose the face that, in the opinion
of the entrant, is the most handsome. A more sophisticated contest
entrant, wishing to maximize the chances of winning a prize, would
think about what the majority perception of attractive is, and then
make a selection based on some inference from his knowledge of public
perceptions. This can be carried one step further to take into account
the fact that other entrants would each have their own opinion of
what public perceptions are. Thus the strategy can be extended to
the next order and the next and so on, at each level attempting to
predict the eventual outcome of the process based on the reasoning
of other rational agents. Keynes believed that similar behavior was
at work within the stock market. This would have people pricing shares
not based on what they think their fundamental value is, but rather
on what they think everyone else thinks their value is, or what everybody
else would predict the average assessment of value to be. 
\end{doublespace}
\item \label{Nash-Equilibrium}\href{https://en.wikipedia.org/wiki/Nash_equilibrium}{Nash Equilibrium, Wikipedia Link}In
game theory, the Nash equilibrium, named after American mathematician
John Forbes Nash Jr., is a solution concept of a non-cooperative game
involving two or more players in which each player is assumed to know
the equilibrium strategies of the other players, and no player has
anything to gain by changing only their own strategy.
\item \label{Guess-Game-Classroom}\href{https://en.wikipedia.org/wiki/Guess_2/3_of_the_average\#Experimental_results}{Experimental Results of Guess 2/3 of the average, Wikipedia Link}We
have tried more than ten trials of this game as a classroom experiment
with students of economics and finance. The number of participants
varied from 25 to around 60. When the number of participants were
large, the students were asked to provide their answers in teams of
three or four so that the total number of answers were usually between
15 to 20. This was done only to reduce the overhead later in finalizing
the results. The students played the game to obtain a certain advantage
in their final score for the course. The winning guess has usually
been between 20 to 30. An interesting note is that when the team size
is more than one, it leads to a lower guess.
\begin{doublespace}
\item \label{MSCI}\href{https://en.wikipedia.org/wiki/MSCI}{MSCI, Wikipedia Link}
MSCI Inc. (formerly Morgan Stanley Capital International and MSCI
Barra), is a Global provider of equity, fixed income, hedge fund stock
market indexes, and multi-asset portfolio analysis tools. It publishes
the MSCI BRIC, MSCI World and MSCI EAFE Indexes. The company is currently
headquartered at 7 World Trade Center in Manhattan, New York City,
U.S.
\item \label{Factor Analysis}\href{https://en.wikipedia.org/wiki/Factor_analysis}{Factor Analysis, Wikipedia Link}
Factor analysis is a statistical method used to describe variability
among observed, correlated variables in terms of a potentially lower
number of unobserved variables called factors. For example, it is
possible that variations in six observed variables mainly reflect
the variations in two unobserved (underlying) variables. Factor analysis
searches for such joint variations in response to unobserved latent
variables. The observed variables are modelled as linear combinations
of the potential factors, plus \textquotedbl error\textquotedbl{}
terms. Factor analysis aims to find independent latent variables.
The theory behind factor analytic methods is that the information
gained about the interdependencies between observed variables can
be used later to reduce the set of variables in a dataset.
\item \label{Equity Swap}\href{https://en.wikipedia.org/wiki/Equity_swap}{Equity Swap, Wikipedia Link}
An equity swap is a financial derivative contract (a swap) where a
set of future cash flows are agreed to be exchanged between two counterparties
at set dates in the future. The two cash flows are usually referred
to as \textquotedbl legs\textquotedbl{} of the swap; one of these
\textquotedbl legs\textquotedbl{} is usually pegged to a floating
rate such as LIBOR. This leg is also commonly referred to as the \textquotedbl floating
leg\textquotedbl . The other leg of the swap is based on the performance
of either a share of stock or a stock market index. This leg is commonly
referred to as the \textquotedbl equity leg\textquotedbl . Most
equity swaps involve a floating leg vs. an equity leg, although some
exist with two equity legs. An equity swap involves a notional principal,
a specified duration and predetermined payment intervals. The term
\textquotedbl tenor\textquotedbl{} may refer either to the duration
or the coupon frequency. Equity swaps are typically traded by Delta
One trading desks.
\end{doublespace}
\end{enumerate}
\begin{doublespace}

\section{References }
\end{doublespace}
\begin{enumerate}
\begin{doublespace}
\item Adcock, C. J. (1997). Sample size determination: a review. Journal
of the Royal Statistical Society: Series D (The Statistician), 46(2),
261-283.
\item Aked, M., \& Moroz, M. (2015). The Market Impact of Passive Trading.
Journal of Trading, 10(3), 5-12.
\end{doublespace}
\item Alberts, B., Johnson, A., Lewis, J., Raff, M., Roberts, K., \& Walter,
P. (2002). Cell junctions. In Molecular Biology of the Cell. 4th edition.
Garland Science.
\begin{doublespace}
\item Alexander, C. (1999). Optimal hedging using cointegration. Philosophical
Transactions of the Royal Society of London. Series A: Mathematical,
Physical and Engineering Sciences, 357(1758), 2039–2058.
\item Alexander, C., \& Dimitriu, A. (2005a). Indexing, cointegration and
equity market regimes. International Journal of Finance \& Economics,
10(3), 213-231.
\item Alexander, C., \& Dimitriu, A. (2005b). Indexing and statistical arbitrage.
The Journal of Portfolio Management, 31(2), 50-63.
\item Almgren, R., \& Neil, C. (2001). Optimal execution of portfolio transactions.
Journal of Risk, 3, 5–40. 
\item Amaya, D., Christoffersen, P., Jacobs, K., \& Vasquez, A. (2015).
Does realized skewness predict the cross-section of equity returns?.
Journal of Financial Economics, 118(1), 135-167.
\item Amihud, Y., \& Mendelson, H. (1989). The Effects of Beta, Bid‐Ask
Spread, Residual Risk, and Size on Stock Returns. The Journal of Finance,
44(2), 479-486.
\item Amihud, Y., Mendelson, H., \& Uno, J. (1999). Number of shareholders
and stock prices: Evidence from Japan. The Journal of finance, 54(3),
1169-1184.
\item An, H., \& Zhang, T. (2013). Stock price synchronicity, crash risk,
and institutional investors. Journal of Corporate Finance, 21, 1-15.
\item Andersen, T. G., \& Bollerslev, T. (1998). Answering the skeptics:
Yes, standard volatility models do provide accurate forecasts. International
economic review, 885-905.
\item Andersen, T. G., Bollerslev, T., Diebold, F. X., \& Labys, P. (2003).
Modeling and forecasting realized volatility. Econometrica, 71(2),
579-625.
\item Andersen, J. V., \& Sornette, D. (2004). Fearless versus fearful speculative
financial bubbles. Physica A: Statistical Mechanics and its Applications,
337(3-4), 565-585.
\item Ang, A., Hodrick, R. J., Xing, Y., \& Zhang, X. (2006). The cross‐section
of volatility and expected returns. The Journal of Finance, 61(1),
259-299.
\item Angeletos, G. M., Hellwig, C., \& Pavan, A. (2007). Dynamic global
games of regime change: Learning, multiplicity, and the timing of
attacks. Econometrica, 75(3), 711-756.
\item Badrinath, S. G., \& Chatterjee, S. (1988). On measuring skewness
and elongation in common stock return distributions: The case of the
market index. Journal of Business, 451-472.
\item Badrinath, S. G., \& Chatterjee, S. (1991). A data-analytic look at
skewness and elongation in common-stock-return distributions. Journal
of Business \& Economic Statistics, 9(2), 223-233.
\item Bailey, D. H., Borwein, J. M., de Prado, M. L., \& Zhu, Q. J. (2014).
Pseudo-mathematics and financial charlatanism: the effefcts of backtest
overfitting on out-of-sample performance. Notices of the American
Mathematical Society, 61(5), 458-471.
\item Bakshi, G., Kapadia, N., \& Madan, D. (2003). Stock return characteristics,
skew laws, and the differential pricing of individual equity options.
The Review of Financial Studies, 16(1), 101-143.
\item Banerjee, A., Dolado, J. J., Galbraith, J. W., \& Hendry, D. (1993).
Co-integration, error correction, and the econometric analysis of
non-stationary data. OUP Catalogue.
\item Barberis, N., Mukherjee, A., \& Wang, B. (2016). Prospect theory and
stock returns: an empirical test. The Review of Financial Studies,
29(11), 3068-3107.
\item Beasley, J. E., Meade, N., \& Chang, T. J. (2003). An evolutionary
heuristic for the index tracking problem. European Journal of Operational
Research, 148(3), 621-643.
\item Bernstein, P. L. (1987). Liquidity, stock markets, and market makers.
Financial Management, 54-62.
\item Bhide, A. (1993). The hidden costs of stock market liquidity. Journal
of financial economics, 34(1), 31-51.
\item Bilbiie, F. O., Ghironi, F., \& Melitz, M. J. (2012). Endogenous entry,
product variety, and business cycles. Journal of Political Economy,
120(2), 304-345.
\item Bishop, Y. M., Fienberg, S. E., \& Holland, P. W. (2007). Discrete
Multivariate Analysis: Theory and Practice. Springer Science \& Business
Media.
\item Black, F. (1992). Beta and return. Journal of portfolio management,
1.
\item Blackstone Jr, J. H., \& Cox, J. F. (1985). Inventory management techniques.
Journal of Small Business Management (pre-1986), 23(000002), 27.
\item Bloom, S. M., Gouws, F., \& Holmes, D. T. (2000). U.S. Patent No.
6,061,663. Washington, DC: U.S. Patent and Trademark Office.
\item Bloomberg, L. P. (1981). Bloomberg Terminal. New York: Bloomberg LP.
\item Bodie, Z., Kane, A., \& Marcus, A. (2014). Investments (10th global
ed.). Berkshire: McGraw-Hill Education.
\item Bollerslev, T., Chou, R. Y., \& Kroner, K. F. (1992). ARCH modeling
in finance: A review of the theory and empirical evidence. Journal
of econometrics, 52(1-2), 5-59.
\item Bouchaud, J. P., \& Potters, M. (2003). Theory of financial risk and
derivative pricing: from statistical physics to risk management. Cambridge
university press.
\end{doublespace}
\item Büren, C., Frank, B., \& Nagel, R. (2012). A historical note on the
beauty contest (No. 11-2012). Joint discussion paper series in economics.
\begin{doublespace}
\item Buss, D. M. (1985). Human mate selection: Opposites are sometimes
said to attract, but in fact we are likely to marry someone who is
similar to us in almost every variable. American scientist, 73(1),
47-51.
\item Burstein, G. (1999). Macro trading and investment strategies: macroeconomic
arbitrage in global markets (Vol. 3). John Wiley \& Sons.
\item Cai, J. (1994). A Markov model of switching-regime ARCH. Journal of
Business \& Economic Statistics, 12(3), 309-316.
\item Chan, L. K., \& Lakonishok, J. (1993). Institutional trades and intraday
stock price behavior. Journal of Financial Economics, 33(2), 173-199.
\item Chance, D. M., \& Rich, D. R. (1998). The pricing of equity swaps
and swaptions. The Journal of Derivatives, 5(4), 19-31.
\item Chen, J., Hong, H., \& Stein, J. C. (2001). Forecasting crashes: Trading
volume, past returns, and conditional skewness in stock prices. Journal
of financial Economics, 61(3), 345-381.
\item Chen, A. S., Leung, M. T., \& Daouk, H. (2003). Application of neural
networks to an emerging financial market: forecasting and trading
the Taiwan Stock Index. Computers \& Operations Research, 30(6), 901-923.
\item Cheng, Y., Liu, M. H., \& Qian, J. (2006). Buy-side analysts, sell-side
analysts, and investment decisions of money managers. Journal of Financial
and Quantitative Analysis, 41(1), 51-83.
\item Chiarella, C., \& He, X. (2001). Asset price and wealth dynamics under
heterogeneous expectations. Quantitative Finance, 1(5), 509-526.
\item Church, G. M., Gao, Y., \& Kosuri, S. (2012). Next-generation digital
information storage in DNA. Science, 1226355.
\item Chow, T. M., Hsu, J., Kalesnik, V., \& Little, B. (2011). A survey
of alternative equity index strategies. Financial Analysts Journal,
67(5), 37-57.
\item Christoffersen, P. F. (2012). Elements of financial risk management.
Academic Press.
\item Chung, Y. P. (1991). A transactions data test of stock index futures
market efficiency and index arbitrage profitability. The Journal of
Finance, 46(5), 1791-1809.
\end{doublespace}
\item Copeland, L. S. (2008). Exchange rates and international finance.
Pearson Education.
\begin{doublespace}
\item Corrado, C. J., \& Su, T. (1996). Skewness and kurtosis in S\&P 500
index returns implied by option prices. Journal of Financial research,
19(2), 175-192.
\item Costello, A. B., \& Osborne, J. W. (2005). Best practices in exploratory
factor analysis: Four recommendations for getting the most from your
analysis. Practical assessment, research \& evaluation, 10(7), 1-9.
\item Cremers, M., \& Weinbaum, D. (2010). Deviations from put-call parity
and stock return predictability. Journal of Financial and Quantitative
Analysis, 45(2), 335-367.
\item Datar, V. T., Naik, N. Y., \& Radcliffe, R. (1998). Liquidity and
stock returns: An alternative test. Journal of Financial Markets,
1(2), 203-219.
\item Dose, C., \& Cincotti, S. (2005). Clustering of financial time series
with application to index and enhanced index tracking portfolio. Physica
A: Statistical Mechanics and its Applications, 355(1), 145-151.
\item Driessen, J., Maenhout, P. J., \& Vilkov, G. (2009). The price of
correlation risk: Evidence from equity options. The Journal of Finance,
64(3), 1377-1406.
\item Dwyer Jr, G. P., Locke, P., \& Yu, W. (1996). Index arbitrage and
nonlinear dynamics between the S\&P 500 futures and cash. The Review
of Financial Studies, 9(1), 301-332.
\item Eisenbeis, R. A. (1977). Pitfalls in the application of discriminant
analysis in business, finance, and economics. The Journal of Finance,
32(3), 875-900.
\end{doublespace}
\item Elena, S. F., \& Lenski, R. E. (2003). Microbial genetics: evolution
experiments with microorganisms: the dynamics and genetic bases of
adaptation. Nature Reviews Genetics, 4(6), 457.
\begin{doublespace}
\item Elton, E. J., Gruber, M. J., Brown, S. J., \& Goetzmann, W. N. (2009).
Modern portfolio theory and investment analysis. John Wiley \& Sons.
\item Engle, R. F., \& Clive, W. J. G. (1987). Co-integration and error
correction: representation, estimation, \& testing. Econometrica:
Journal of the Econometric Society, 251–276. 
\item Engle, R. F., \& Ng, V. K. (1993). Measuring and testing the impact
of news on volatility. The journal of finance, 48(5), 1749-1778.
\item Fama, E. F. (1970). Efficient capital markets: A review of theory
and empirical work. The journal of Finance, 25(2), 383-417.
\item Finucane, T. J. (1991). Put-call parity and expected returns. Journal
of Financial and Quantitative Analysis, 26(4), 445-457.
\item Fletcher, J. (2000). On the conditional relationship between beta
and return in international stock returns. International Review of
Financial Analysis, 9(3), 235-245.
\item Franci, F., Marschinski, R., \& Matassini, L. (2001). Learning the
optimal trading strategy. Physica A: Statistical Mechanics and its
Applications, 294(1-2), 213-225.
\item French, K. R., Schwert, G. W., \& Stambaugh, R. F. (1987). Expected
stock returns and volatility. Journal of financial Economics, 19(1),
3-29.
\item Fung, J. K., Lau, F., \& Tse, Y. (2015). The impact of sampling frequency
on intraday correlation and lead–lag relationships between index futures
and individual stocks. Journal of Futures Markets, 35(10), 939-952.
\end{doublespace}
\item Gans, J. S., \& Shepherd, G. B. (1994). How are the mighty fallen:
Rejected classic articles by leading economists. Journal of Economic
Perspectives, 8(1), 165-179.
\begin{doublespace}
\item Gasiorowski, M. J. (1995). Economic crisis and political regime change:
An event history analysis. American political science review, 89(4),
882-897.
\item Gay, G., Venkateswaran, A., Kolb, R. W., \& Overdahl, J. A. (2008).
The pricing and valuation of swaps. Financial Derivatives: Pricing
and Risk Management, 405-422.
\item Glosten, L. R., Jagannathan, R., \& Runkle, D. E. (1993). On the relation
between the expected value and the volatility of the nominal excess
return on stocks. The journal of finance, 48(5), 1779-1801.
\item Gompers, P. A., \& Metrick, A. (2001). Institutional investors and
equity prices. The quarterly journal of Economics, 116(1), 229-259.
\item Granger, C. W. (1981). Some properties of time series data and their
use in econometric model specification. Journal of econometrics, 16(1),
121-130.
\item Groysberg, B., Healy, P., \& Chapman, C. (2008). Buy-side vs. sell-side
analysts’ earnings forecasts. Financial Analysts Journal, 64(4), 25-39.
\item Guastaroba, G., \& Speranza, M. G. (2012). Kernel search: An application
to the index tracking problem. European Journal of Operational Research,
217(1), 54-68.
\item Gujarati, D. N. (2009). Basic econometrics. Tata McGraw-Hill Education.
\item Hair, J. F., Black, W. C., Babin, B. J., Anderson, R. E., \& Tatham,
R. L. (1998). Multivariate data analysis (Vol. 5, No. 3, pp. 207-219).
Upper Saddle River, NJ: Prentice hall.
\item Hameed, A., Kang, W., \& Viswanathan, S. (2010). Stock market declines
and liquidity. The Journal of Finance, 65(1), 257-293.
\item Hamilton, J. D. (1994). Time series analysis (Vol. 2). Princeton:
Princeton university press.
\item Harris, R. I. (1995). Using cointegration analysis in econometric
modelling.
\item Hassan, M. R., \& Nath, B. (2005). Stock market forecasting using
hidden Markov model: a new approach. In Intelligent Systems Design
and Applications, 2005. ISDA'05. Proceedings. 5th International Conference
on (pp. 192-196). IEEE.
\item Hassan, M. R. (2009). A combination of hidden Markov model and fuzzy
model for stock market forecasting. Neurocomputing, 72(16-18), 3439-3446.
\item Hau, H., Massa, M., \& Peress, J. (2009). Do demand curves for currencies
slope down? Evidence from the MSCI global index change. The Review
of Financial Studies, 23(4), 1681-1717.
\end{doublespace}
\item Heller, J. (1999). Catch-22: a novel (Vol. 4). Simon and Schuster.
\begin{doublespace}
\item Hong, H., Li, W., Ni, S. X., Scheinkman, J. A., \& Yan, P. (2015).
Days to cover and stock returns (No. w21166). National Bureau of Economic
Research.
\item Hott, C. (2009). Herding behavior in asset markets. Journal of Financial
Stability, 5(1), 35-56.
\item Hull, J. C., \& Basu, S. (2016). Options, futures, and other derivatives.
Pearson Education India.
\item Isakov, D. (1999). Is beta still alive? Conclusive evidence from the
Swiss stock market. The European Journal of Finance, 5(3), 202-212.
\item Ismail, S. (2014). Exponential Organizations: Why new organizations
are ten times better, faster, and cheaper than yours (and what to
do about it). Diversion Books.
\end{doublespace}
\item Jacobs, B. I., \& Levy, K. N. (1993). Long/short equity investing.
Journal of Portfolio Management, 20(1), 52.
\begin{doublespace}
\item Jolliffe, I. T. (1986). Principal component analysis and factor analysis.
In Principal component analysis (pp. 115-128). Springer, New York,
NY.
\item Kashyap, R. (2014a). Dynamic Multi-Factor Bid–Offer Adjustment Model.
The Journal of Trading, 9(3), 42-55. 
\item Kashyap, R. (2014b). The Circle of Investment. International Journal
of Economics and Finance, 6(5), 244-263. 
\item Kashyap, R. (2015). A Tale of Two Consequences. The Journal of Trading,
10(4), 51-95. 
\item Kashyap, R. (2016a). Hong Kong - Shanghai Connect / Hong Kong - Beijing
Disconnect (?), Scaling the Great Wall of Chinese Securities Trading
Costs. The Journal of Trading, 11(3), 81-134. 
\item Kashyap, R. (2016b). David vs Goliath (You against the Markets), A
Dynamic Programming Approach to Separate the Impact and Timing of
Trading Costs. arXiv preprint arXiv:1603.00984.
\item Kashyap, R (2017). Artificial Intelligence: A Child's Play. Working
Paper.
\end{doublespace}
\item Kashyap, R (2018a). Beating Benchmarks by Building Bouncy Baskets.
Working Paper.
\begin{doublespace}
\item Kashyap, R (2018b). Seven Survival Senses: Evolutionary Training makes
Discerning Differences more Natural than Spotting Similarities. Working
Paper.
\item Kat, H. M. (2001). Structured equity derivatives. England: John Wiley
\& Sons.
\end{doublespace}
\item Ke, Q., Ferrara, E., Radicchi, F., \& Flammini, A. (2015). Defining
and identifying Sleeping Beauties in science. Proceedings of the National
Academy of Sciences, 112(24), 7426-7431.
\item Keynes, J. M. (2018). The general theory of employment, interest,
and money. Springer.
\begin{doublespace}
\item Kijima, M., \& Muromachi, Y. (2001). Pricing equity swaps in a stochastic
interest rate economy. The Journal of Derivatives, 8(4), 19-35.
\item Klemkosky, R. C., \& Resnick, B. G. (1979). Put‐Call Parity and Market
Efficiency. The Journal of Finance, 34(5), 1141-1155.
\item Kostovetsky, L. (2003). Index mutual funds and exchange-traded funds.
Journal of Portfolio Management, 29(4), 80-92.
\item Kosuri, S., \& Church, G. M. (2014). Large-scale de novo DNA synthesis:
technologies and applications. Nature methods, 11(5), 499.
\item Krejcie, R. V., \& Morgan, D. W. (1970). Determining sample size for
research activities. Educational and psychological measurement, 30(3),
607-610.
\item Kumar, P., \& Seppi, D. J. (1994). Information and index arbitrage.
Journal of Business, 481-509.
\item Lakonishok, J., Shleifer, A., \& Vishny, R. W. (1992). The impact
of institutional trading on stock prices. Journal of financial economics,
32(1), 23-43.
\item Lancaster, K. (1990). The economics of product variety: A survey.
Marketing science, 9(3), 189-206.
\item Lancioni, R. A., \& Howard, K. (1978). Inventory management techniques.
International Journal of Physical Distribution \& Materials Management,
8(8), 385-428.
\item Lenth, R. V. (2001). Some practical guidelines for effective sample
size determination. The American Statistician, 55(3), 187-193.
\item Levy, M., \& Levy, H. (1996). The danger of assuming homogeneous expectations.
Financial Analysts Journal, 52(3), 65-70.
\item Litterman, R. B. (1983). A random walk, Markov model for the distribution
of time series. Journal of Business \& Economic Statistics, 1(2),
169-173.
\item Ludvigson, S. C., \& Ng, S. (2007). The empirical risk–return relation:
A factor analysis approach. Journal of Financial Economics, 83(1),
171-222.
\item Lutz, J. F., Ouchi, M., Liu, D. R., \& Sawamoto, M. (2013). Sequence-controlled
polymers. Science, 341(6146), 1238149.
\item Maddala, G. S., \& Lahiri, K. (1992). Introduction to econometrics
(Vol. 2). New York: Macmillan.
\item Mantzicopoulos, P., \& Patrick, H. (2010). “The seesaw is a machine
that goes up and down”: Young children's narrative responses to science-related
informational text. Early Education and Development, 21(3), 412-444.
\item Marshall, C. M. (2009). Dispersion trading: Empirical evidence from
US options markets. Global Finance Journal, 20(3), 289-301.
\item MacCallum, R. C., Widaman, K. F., Zhang, S., \& Hong, S. (1999). Sample
size in factor analysis. Psychological methods, 4(1), 84.
\item Meissner, G. (2016). Correlation Trading Strategies–Opportunities
and Limitations. The Journal of Trading, 11(4), 14-32.
\item Merkley, K., Michaely, R., \& Pacelli, J. (2017). Does the scope of
the sell‐side analyst industry matter? An examination of bias, accuracy,
and information content of analyst reports. The Journal of Finance,
72(3), 1285-1334.
\item Miao, G. J. (2014). High frequency and dynamic pairs trading based
on statistical arbitrage using a two-stage correlation and cointegration
approach. International Journal of Economics and Finance, 6(3), 96.
\end{doublespace}
\item Nagel, R. (1995). Unraveling in guessing games: An experimental study.
The American Economic Review, 85(5), 1313-1326.
\item Nagel, R., Bühren, C., \& Frank, B. (2017). Inspired and inspiring:
Hervé Moulin and the discovery of the beauty contest game. Mathematical
Social Sciences, 90, 191-207.
\item Nash, J. (1951). Non-cooperative games. Annals of mathematics, 286-295.
\begin{doublespace}
\item Nandan, T., Agrawal, P., \& Jindal, S. (2015). An Empirical Investigation
of Mispricing in Stock Futures at the National Stock Exchange. Indian
Journal of Finance, 9(9), 23-35.
\item Natenberg, S. (1994). Option volatility \& pricing: advanced trading
strategies and techniques. McGraw Hill Professional.
\end{doublespace}
\item Nei, M. (2013). Mutation-driven evolution. OUP Oxford.
\begin{doublespace}
\item Nekrasov, V. (2014). Knowledge Rather Than Hope: a Book for Retail
Investors and Mathematical Finance Students. Vasily Nekrasov.
\item Nelson, D. B. (1991). Conditional heteroskedasticity in asset returns:
A new approach. Econometrica: Journal of the Econometric Society,
347-370.
\item Ng, V. K., Engle, R. F., \& Rothschild, M. (1992). A multi-dynamic-factor
model for stock returns.
\end{doublespace}
\item Osborne, M. J., \& Rubinstein, A. (1994). A course in game theory.
MIT press.
\begin{doublespace}
\item Pardo, R. (2011). The evaluation and optimization of trading strategies
(Vol. 314). John Wiley \& Sons.
\item Pástor, Ľ., \& Stambaugh, R. F. (2003). Liquidity risk and expected
stock returns. Journal of Political economy, 111(3), 642-685.
\item Perold., R. F. (1998). The implementation shortfall: Paper versus
reality. Streetwise: the best of the Journal of Portfolio Management,
106. 
\end{doublespace}
\item Persky, J. (1995). The ethology of homo economicus. Journal of Economic
Perspectives, 9(2), 221-231.
\begin{doublespace}
\item Plomin, R., \& Daniels, D. (1987). Why are children in the same family
so different from one another?. Behavioral and brain Sciences, 10(1),
1-16.
\item Randall, T., \& Ulrich, K. (2001). Product variety, supply chain structure,
and firm performance: Analysis of the US bicycle industry. Management
Science, 47(12), 1588-1604.
\item Reynolds, R. (1992). Super heroes: A modern mythology. Univ. Press
of Mississippi.
\end{doublespace}
\item Rosenberg, S. M. (1997). Mutation for survival. Current opinion in
genetics \& development, 7(6), 829-834.
\begin{doublespace}
\item Roll, R., \& Ross, S. A. (1980). An empirical investigation of the
arbitrage pricing theory. The Journal of Finance, 35(5), 1073-1103.
\item Ross, S. A. (1976). The arbitrage theory of capital asset pricing.
Journal of Economic Theory, 13(3), 341-360.
\item Roy, R. K., Meszynska, A., Laure, C., Charles, L., Verchin, C., \&
Lutz, J. F. (2015). Design and synthesis of digitally encoded polymers
that can be decoded and erased. Nature communications, 6, 7237.
\item Scherbina, A., \& Schlusche, B. (2014). Asset price bubbles: a survey.
Quantitative Finance, 14(4), 589-604.
\item Schipper, K. (1991). Analysts' forecasts. Accounting horizons, 5(4),
105.
\item Schwert, G. W. (1989). Why does stock market volatility change over
time?. The journal of finance, 44(5), 1115-1153.
\item Sharpe, W. F. (1964). Capital asset prices: A theory of market equilibrium
under conditions of risk. The journal of finance, 19(3), 425-442.
\item Shlens, J. (2005). A tutorial on principal component analysis. Systems
Neurobiology Laboratory, University of California at San Diego.
\item Siegel, J. J. (2003). What is an asset price bubble? An operational
definition. European financial management, 9(1), 11-24.
\item Smithson, C. W. (1998). Managing financial risk. McGraw-Hill.
\item Spence, M. (1976). Product selection, fixed costs, and monopolistic
competition. The Review of economic studies, 43(2), 217-235.
\item Srinivasan, S., Pauwels, K., Silva-Risso, J., \& Hanssens, D. M. (2009).
Product innovations, advertising, and stock returns. Journal of Marketing,
73(1), 24-43.
\item Stocker, T. F. (1998). The seesaw effect. Science, 282(5386), 61-62.
\item Stöckl, T., Huber, J., \& Kirchler, M. (2010). Bubble measures in
experimental asset markets. Experimental Economics, 13(3), 284-298.
\item Stoyan, S. J., \& Kwon, R. H. (2010). A two-stage stochastic mixed-integer
programming approach to the index tracking problem. Optimization and
Engineering, 11(2), 247-275.
\item Taleb, N. N. (2007). The black swan: The impact of the highly improbable
(Vol. 2). Random house.
\end{doublespace}
\item Thaler, R. H. (2000). From homo economicus to homo sapiens. Journal
of economic perspectives, 14(1), 133-141.
\begin{doublespace}
\item Tuckman, B., \& Serrat, A. (2011). Fixed income securities: tools
for today's markets (Vol. 626). John Wiley \& Sons.
\item Turner, C. M., Startz, R., \& Nelson, C. R. (1989). A Markov model
of heteroskedasticity, risk, and learning in the stock market. Journal
of Financial Economics, 25(1), 3-22.
\item Valsiner, J. (2007). Culture in minds and societies: Foundations of
cultural psychology. Psychol. Stud.(September 2009), 54, 238-239.
\item Verbeek, M. (2008). A guide to modern econometrics. John Wiley \&
Sons.
\item Wade, R. H. (2008). Financial regime change?. New Left Review, 53,
5-21.
\item Wang, M. C., \& Liao, S. L. (2003). Pricing models of equity swaps.
Journal of Futures Markets, 23(8), 751-772.
\item Williams, P. A., Moyes, G. D., \& Park, K. (1996). Factors affecting
earnings forecast revisions for the buy-side and sell-side analyst.
Accounting Horizons, 10(3), 112.
\end{doublespace}
\item Wilke, C. O., Wang, J. L., Ofria, C., Lenski, R. E., \& Adami, C.
(2001). Evolution of digital organisms at high mutation rates leads
to survival of the flattest. Nature, 412(6844), 331.
\begin{doublespace}
\item Window, C. (2010). What is MATLAB?.
\item Wipplinger, E. (2007). Philippe Jorion: Value at Risk-The New Benchmark
for Managing Financial Risk. Financial Markets and Portfolio Management,
21(3), 397.
\item Wu, T. P., \& Chen, S. N. (2007). Equity swaps in a LIBOR market model.
Journal of Futures Markets, 27(9), 893-920.
\item Xu, Y., \& Malkiel, B. G. (2003). Investigating the Behavior of Idiosyncratic
Volatility. The Journal of Business, 76(4), 613-645.
\end{doublespace}
\end{enumerate}

\end{document}